\documentclass[nofootinbib,superscriptaddress,onecolumn,preprintnumbers]{revtex4}
\usepackage{graphicx}
\usepackage{amsmath,amssymb}
\usepackage[normalem]{ulem}
\renewcommand{\eqref}[1]{Eq.(\ref{#1})}

\usepackage{color}

\begin{document}
\title{Modifying $\Lambda$CDM dynamics via out-of-equilibrium axions: \\ reconciling SH0ES and DESI $H_0$ values}

\author{Giovanni Montani}
\email{giovanni.montani@enea.it}
\affiliation{Nuclear Department, ENEA - C. R. Frascati, Via E. Fermi 45, 00044 Frascati, Italy}
\affiliation{Physics Department, ``Sapienza'' University of Rome,  P.le Aldo Moro 5, 00185 Roma, Italy}

\author{Luis A. Escamilla}
\affiliation{Department of Physics, Istanbul Technical University, 34469 Maslak, Istanbul, Turkey}

\author{Nakia Carlevaro}
\affiliation{Nuclear Department, ENEA - C. R. Frascati, Via E. Fermi 45, 00044 Frascati, Italy}

\author{Francesco Cianfrani}
\affiliation{Nuclear Department, ENEA - C. R. Frascati, Via E. Fermi 45, 00044 Frascati, Italy}

\begin{abstract}
We investigate late-Universe dynamics in which the dark matter component is described by axion particles. The proposed framework departs from the standard $\Lambda$CDM paradigm due to a small fraction of axions driving the system away from thermal equilibrium. We analyze the evolution of the axion energy density using both a kinetic and a classical field approach, yielding an identical macroscopic evolution equation for the dark matter density. 
We emphasize that the BGK parameter is introduced phenomenologically at the kinetic level and this does not supply an independent microscopic derivation. The present work therefore explores the phenomenological consequences of late-time, out‑of‑equilibrium axion production rather than claiming a completed microphysical model. The resulting scenario modifies $\Lambda$CDM dynamics in the late Universe (specifically at $z \lesssim 1$), while asymptotically recovering the standard baseline at earlier cosmic epochs. We compare the theoretical predictions of our formulation against a comprehensive suite of late-Universe datasets. Our statistical analysis reveals that when the SH0ES local calibration is included, the collisional axion model becomes significantly favored over $\Lambda$CDM, yielding a best-fit Hubble constant of $H_0 \simeq 73~{\rm km\,s^{-1}\,Mpc^{-1}}$. Ultimately, this cosmological scenario successfully accommodates local distance-ladder measurements while maintaining excellent agreement with Baryon Acoustic Oscillation data from the DESI Collaboration.
\end{abstract}

\maketitle

\section{Introduction}
The capability of the $\Lambda$CDM model \cite{Weinberg2008,KolbTurner1990} to reliably reproduce the observed Universe---across both its early and late stages of evolution---has been significantly challenged by the persistent tension in the determination of the Hubble constant~\cite{Verde:2019ivm,DiValentino:2020zio,DiValentino:2021izs,Perivolaropoulos:2021jda,Schoneberg:2021qvd,Shah:2021onj,Abdalla:2022yfr,DiValentino:2022fjm,Kamionkowski:2022pkx,Hu:2023jqc,Giare:2023xoc,Vagnozzi:2023nrq,Verde:2023lmm,DiValentino:2024yew,Perivolaropoulos:2024yxv,Akarsu:2024qiq,CosmoVerseNetwork:2025alb}. This discrepancy arises between local distance-ladder measurements from the SH0ES Collaboration, which favor a higher value of $H_0 \simeq 73~{\rm km\,s^{-1}\,Mpc^{-1}}$~\cite{Riess:2021jrx,Breuval:2024lsv}, and early-Universe inferences within the $\Lambda$CDM framework from cosmic microwave background (CMB) observations, which prefer a lower value of $H_0 \simeq 67~{\rm km\,s^{-1}\,Mpc^{-1}}$~\cite{Planck:2018vyg,SPT-3G:2025bzu}. Crucially, recent comparisons between the updated SPT-3G $\Lambda$CDM inference and the H0DN ``Local Distance Network'' value of $H_0=73.50\pm0.81~{\rm km\,s^{-1}\,Mpc^{-1}}$ have revealed a discrepancy of $\simeq 7.1\sigma$~\cite{SPT-3G:2025bzu,H0DN:2025lyy}, exacerbating this tension to unprecedented statistical significance. 

Over the last decade, extensive effort has been devoted to resolving the persistent Hubble tension through modifications targeting different epochs and sectors of the cosmological framework \cite{DiValentino:2021izs, Kamionkowski23, 2025CQGra..42x5002S}. Proposed solutions are commonly grouped into two broad categories based on \emph{when} they alter the expansion history: \emph{early-time modifications}, which modify the expansion rate or energy content prior to recombination (such as Early Dark Energy, EDE~\cite{Poulin:2018cxd,Karwal:2016vyq,Hill:2020osr,Ivanov:2020ril,Sakstein:2019fmf,Niedermann:2019olb,Niedermann:2020dwg,Poulin:2023lkg,Smith:2025grk,Poulin:2025nfb,SPT-3G:2025vyw}), and \emph{late-time modifications}, which deform the post-recombination expansion history while preserving the high-redshift successes of the standard paradigm (such as Interacting Dark Energy, IDE~\cite{Caprini:2016qxs,Nunes:2016dlj,Kumar:2017dnp,DiValentino:2017iww,Yang:2017ccc,Costa:2018aoy,vonMarttens:2018iav,Yang:2018euj,Yang:2018uae,Pan:2019gop,Kumar:2019wfs,DiValentino:2019jae,DiValentino:2019ffd,DiValentino:2020kpf,Gomez-Valent:2020mqn,Lucca:2020zjb,Pan:2020zza,Gao:2021xnk,Kumar:2021eev,Yang:2021hxg,Nunes:2022bhn,Bernui:2023byc,Escamilla:2023shf,Giare:2024smz,Li:2024qso,Sabogal:2025mkp,Silva:2025hxw,Yang:2025uyv,vanderWesthuizen:2025rip, Li:2025owk, Li:2026xaz}). Complementarily, departures can be classified by \emph{which sector} is modified, spanning non-standard dark matter properties~\cite{Feng:2010gw,Dodelson:1993je,Joyce:2014kja,Abazajian:2012ys}, exotic late-time dark energy dynamics~\cite{Copeland:2006wr,Benisty:2021gde,Benisty:2020otr,Bamba:2012cp, Wolf:2024eph}, and large-scale modifications of gravity~\cite{Clifton:2011jh,CANTATA:2021ktz,Bahamonde:2021gfp,AlvesBatista:2021gzc,Addazi:2021xuf,Capozziello:2011et}. Because the direct detection of particle dark matter remains elusive~\cite{Baudis:2016qwx,Bertone:2004pz}, the possibility remains open that this emerging phenomenology reflects novel dark sector or gravitational physics.

From a phenomenological perspective, a successful late-Universe solution typically requires the emergence of an effective phantom energy density relative to the standard baseline~\cite{2021MNRAS.505.3866E,2003RvMP...75..559P}, which is hypothesized to be an effective signature of an underlying microphysical mechanism~\cite{2025PDU....4801847M}. Parallel evidence for modifying $\Lambda$CDM at late times stems from the analysis of binned Type Ia Supernova (SN Ia) samples~\cite{Dainotti2021apj-powerlaw, 2025arXiv250111772D, Dainottigalaxies10010024}. These studies outline a decaying, power-law behavior of the localized Hubble constant across distinct redshift bins, which can be naturally extrapolated to the Planck-CMB value within uncertainties. This trend has been physically explored using diagnostic frameworks like the effective running ``Hubble constant'' \cite{2024arXiv240801410S, schiavone_mnras, montaniEntropy}, evolutionary dark energy schemes \cite{fazzari2025, navone2025}, or models exhibiting redshift-dependent $H_0$ decay \cite{kazantzidis, Krishnan:2020vaf}.

This backdrop of theoretical and localized indications has been profoundly sharpened by recent findings from the Dark Energy Spectroscopic Instrument (DESI) Collaboration~\cite{DESI:2024mwx,DESI:2025zgx}. The DESI analysis demonstrates that an evolving dark energy term provides a superior fit to observations than a pure cosmological constant, indicating that $\Lambda$CDM may not be fully predictive in the low-redshift regime. Specifically, the Chevallier-Polarski-Linder (CPL) parametrization~\cite{Chevallier2001,Linder2003} yields a significantly better fit to Baryon Acoustic Oscillation (BAO) data. This revelation has catalyzed a comprehensive body of systematic explorations, confirming that extensions like Dynamical Dark Energy (DDE) substantially enhance the joint consistency of BAO and supernova datasets relative to $\Lambda$CDM~\cite{Colgain:2024xqj,Cortes:2024lgw,Shlivko:2024llw,Luongo:2024fww,Yin:2024hba,Gialamas:2024lyw,Dinda:2024kjf,Najafi:2024qzm,Wang:2024dka,Ye:2024ywg,Tada:2024znt,Carloni:2024zpl,Chan-GyungPark:2024mlx,DESI:2024kob,Bhattacharya:2024hep,Ramadan:2024kmn,Notari:2024rti,Orchard:2024bve,Hernandez-Almada:2024ost,Pourojaghi:2024tmw,Giare:2024gpk,Reboucas:2024smm,Giare:2024ocw,Chan-GyungPark:2024brx,Menci:2024hop,Li:2024qus,Li:2024hrv,Notari:2024zmi,Gao:2024ily,Fikri:2024klc,Jiang:2024xnu,Zheng:2024qzi,Gomez-Valent:2024ejh,RoyChoudhury:2024wri,Li:2025cxn,Lewis:2024cqj,Wolf:2025jlc,Shajib:2025tpd,Giare:2025pzu,Chaussidon:2025npr,Kessler:2025kju,Pang:2025lvh,Roy:2024kni,RoyChoudhury:2025dhe,Paliathanasis:2025cuc,Scherer:2025esj,Giare:2024oil,Liu:2025mub,Teixeira:2025czm,Santos:2025wiv,Specogna:2025guo,Sabogal:2025jbo,Cheng:2025lod,Herold:2025hkb,Cheng:2025hug,Ozulker:2025ehg,Lee:2025pzo,Ormondroyd:2025iaf,Silva:2025twg,Ishak:2025cay,Fazzari:2025lzd,2026JHEAp..5200579O,2026arXiv260104048P,2026PhRvD.113d3545G,2026arXiv260203928B,2026arXiv260110127L,2026arXiv260100650P,2026arXiv260117938S,2026arXiv260105218S,2026arXiv260403756P,2025arXiv250925812P,RoyChoudhury:2025iis,Wolf:2024stt,Wolf:2025acj,Wolf:2025jed,DeSimone:2024lvy,Dainotti:2022bzg,Dainotti:2021pqg,Dainotti:2025qxz}.

In this paper, we propose that the physical process accounting for the Hubble tension (and thus the effective phantom contribution) is a deviation for thermal equilibrium  undergone by a fraction of the axion component of the Universe, here associated with cold dark matter \cite{peacock, OHare:2024nmr, Ringwald_2025}. We analyze this phenomenon using both kinetic and classical scalar field formulations, which yield a consistent equation for the axion energy density perturbation. The central idea is that the thermal equilibrium of the axion component is weakly broken at low redshifts, resulting in a mechanism responsible for axion creation \cite{elizaldeOdintsov}. Consequently, the Hubble parameter approaches the $\Lambda$CDM dynamics, specifically the trajectory dictated by the combination of DESI and Planck data, at sufficiently large redshift values ($z \sim 1$). We compare this model against a comprehensive suite of low-redshift datasets and trace the implications for the mitigation of the Hubble tension.

It is important to note that our discussion below refers to a BGK term with opposite sign relative to the standard collisional term \cite{PhysRev.94.511}, though this interpretation is rigorously applicable only when using redshift as the time variable: in the present epoch, axions exhibit non-equilibrium characteristics, whereas at higher redshift values, thermodynamic equilibrium is completely restored and the cosmological dynamics converge to a standard $\Lambda$CDM model. When following the synchronous time coordinate instead, the non-equilibrium behavior manifests in the axion distribution at low redshift, producing a macroscopic effect equivalent to axion creation (for potential explanations, see the discussion at the conclusion). To strengthen the validity of the macroscopic equation governing axion energy density evolution, we also derive it below using a purely classical field theory approach, which provides additional insight into the phenomenology and dynamics of axion particles.

The BGK term should be viewed as a phenomenological input to the kinetic equation. The presented classical scalar field formulation reproduces the same macroscopic dynamics but does not independently explain the microphysics behind it. Therefore, our results address the observable consequences of a late out‑of‑equilibrium axion component, not a derived particle‑physics mechanism.

The paper is organized as follows. Section \ref{sec1} reviews axion basics (potential, misalignment, energy density scaling) and shows axions behave as pressureless dark matter for typical masses relevant to CDM; section \ref{sec2} models axion distribution as equilibrium plus a perturbation obeying a Boltzmann equation with a BGK-like source term (opposite sign) that yields a modified redshift evolution for the axion density. In subsection \ref{sec2a} the same macroscopic axion-density evolution is derived from a classical field perturbation with an effective damping/source term, showing consistency with the kinetic treatment.
Section \ref{sec3} introduces model parameters, describes priors and datasets (CC, BAO/DESI, Pantheon+, Union3, SH0ES), and explains Bayesian model comparison.
Subsection \ref{sec3a} reports posteriors and model comparison: the present model is preferred when including the SH0ES calibration—then it fits $H_0 = 72.95\pm 0.89~{\rm km\,s^{-1}\,Mpc^{-1}}$ and is strongly favored by Bayesian evidence ($lnB \simeq +5.32$) compared with $\Lambda$CDM. Brief conclusions follow in section \ref{sec4}.

\section{Axion phenomenology}\label{sec1}
The axion field is a scalar boson corresponding to the phase $\theta$ of a complex field, whose modulus represents the Higgs boson of a spontaneous symmetry breaking (SSB) occurring at the energy scale $\sigma_{SSB}$. The potential term describing the axion dynamics takes the form:
\begin{equation}
	V(\theta ) = m_a^2\sigma_{SSB}^2\left( 1 - \cos \theta \right)
	\, , 
	\label{ass6}
\end{equation}
where $m_a$ is the axion rest mass. 

In a flat, isotropic Universe \cite{2020MNRAS.496L..91E} with the line element (adopting $c=1$ and standard notation)
\begin{equation}
	ds^2 = dt^2 - a^2(t) \delta_{ij}dx^idx^j
	\, ,
	\label{ass7}
\end{equation}
($i,j=1,2,3$) where $a(t)$ is the cosmic scale factor in synchronous time, and the axion field near the vacuum state ($\theta \approx 0$) undergoes a regime of small oscillations (misalignment) governed by:
\begin{equation}
	\ddot{\theta} + 3H\dot{\theta} + m_a^2\theta = 0
	\, .
	\label{ass8}
\end{equation}
Here, dots denote differentiation with respect to cosmic time $t$. This evolution is primarily relevant to the late-Universe scenario; for energy scales exceeding the quark-hadron transition, i.e., $\Lambda_{QCD} \simeq 200$~MeV, the axion potential is negligible. A well-known estimate for the present-day axion critical density $\Omega_a^0$ is given by \cite{peacock,KolbTurner1990}:
\begin{equation}
   \Omega_a^0h^2 \sim \frac{10^{-5}\,\text{eV}}{m_a}
	\, , 
	\label{ass9}
\end{equation}
where $h \equiv H_0/100$~km\,s$^{-1}$\,Mpc$^{-1}$ and $H_0$ is the Hubble constant. 

Consequently, for the axion to be a viable dark matter candidate (i.e., $\Omega_a \sim 1$), its rest mass must be of $10^{-5}$~eV order. This value is significantly larger than the Hubble mass $m_{H_0} \equiv H_0 \sim 10^{-33}$~eV (in natural units). Under this condition, Eq.~(\ref{ass8}) admits the leading-order solution:
\begin{equation}
	\theta (t) = \frac{\theta_0}{a^{3/2}}\sin (m_a t + \theta^*)
	\, , 
	\label{ass10}
\end{equation}
where $\theta_0$ and $\theta^*$ are integration constants. Averaging over many oscillation periods, we obtain the following expressions for the energy density and pressure, respectively:
\begin{equation}
	\rho_a = \frac{\sigma_{SSB}^2}{2} \left( \langle \dot{\theta}^2\rangle + m_a^2\langle \theta^2\rangle \right) \simeq \frac{\sigma_{SSB}^2 m_a^2 \theta_0^2}{2a^3}
	\, , 
	\label{ass11}
\end{equation}
\begin{equation}
	p_a = \frac{\sigma_{SSB}^2}{2}\left( \langle \dot{\theta}^2\rangle - m_a^2\langle \theta^2\rangle \right) \simeq 0
	\, .
	\label{ass12}
\end{equation}
Thus, axions behave as a pressureless dark matter component, effectively acting as a dust fluid within the Hubble flow. From a classical scalar field perspective, the axion occupation number is exceptionally high, allowing them to be statistically represented as a Bose-Einstein condensate \cite{PhysRevLett.103.111301}. This justifies the cold nature of the dark matter candidate despite the small value of $m_a$.

The post-inflationary symmetry breaking scenario provides random $\theta_0$ in different Universe patches. Including a more reliable numerical computation of axion mass and Universe thermal history, the axion misalignment gives an average critical density \cite{OHare:2024nmr,Ringwald_2025}
\begin{equation}
\Omega_a^0h^2 = 0.12 \left(\frac{30\,\mu\text{eV}}{m_a}\right)^{7/6} 
\end{equation}
in good agreement with the estimate in \eqref{ass9}. However, the quantitative role of topological defects in this scenario is still uncertain \cite{10.21468/SciPostPhys.10.2.050,Dine:2020pds,Buschmann:2021sdq}.

\section{Kinetic approach}\label{sec2}
We now formulate a statistical approach by postulating that a small fraction of the cosmological axion component deviates from thermal equilibrium. Consequently, the axion distribution function, $f^a(t,p)$ (where $p$ denotes the spatial momentum), can be decomposed as $f^a = f_0^a + \delta f^a$. Here, $f_0^a$ describes the dominant equilibrium component satisfying the Vlasov equation \cite{KolbTurner1990, montani-primordialcosmology}:
\begin{equation}
	\partial_t f_0^a + Hp \partial_p f_0^a = 0
	\, , 
	\label{ax1}
\end{equation}
while the evolution of the perturbation $\delta f^a$ is governed by a homogeneous Boltzmann equation with a modified BGK relaxation term \cite{PhysRev.94.511}, which breaks the equilibrium over time:
\begin{equation}
	\partial_t \delta f^a + Hp \partial_p \delta f^a = \nu \delta f^a
	\, .
	\label{ax2}
\end{equation}
In this context, $\nu$ is opposite to the conventional collision frequency responsible for relaxation to thermal equilibrium and it models the deviation from the equilibrium state due to axion creation (see the discussion in the conclusions).

The axion energy density is provided by the integral:
\begin{equation}
	\rho_a = \frac{1}{(2\pi)^3} \int_0^{\infty} p^2 E_a f^a dp
	\, , 
	\label{ax3}
\end{equation}
where $E_a$ is the axion energy. This expression applies to both the equilibrium phase ($f_0^a \to \rho_a^{eq}$) and the collisional perturbation ($\delta f^a \to \delta \rho_a$). 

Transforming Eqs.~(\ref{ax1}) and (\ref{ax2}) into macroscopic equations by changing the independent variable from cosmic time $t$ to redshift $z$ (using $\dot{(\dots)} = -H(1+z)d(\dots)/dz$)\footnote{We conventionally set the present-day value of the cosmic scale factor to unity.}, we obtain:
\begin{equation}
	\frac{d\rho_a^{eq}}{dz} - \frac{3}{1+z}\rho_a^{eq} = 0
	\, 
	\label{ax4}
\end{equation}
and
\begin{equation}
	\frac{d\delta\rho_a}{dz} - \frac{3}{1+z}\delta\rho_a = -\frac{\nu}{(1+z)H}\delta\rho_a
	\, , 
	\label{ax5}
\end{equation}
respectively. Note that the axion pressure remains negligible in both phases because the available momentum states remain concentrated near the ground state of the Bose-Einstein condensate. Solving Eq.~(\ref{ax4}) and combining it with Eq.~(\ref{ax5}), we derive the following evolution equation for the total axion density $\rho_a \equiv \rho_a^{eq} + \delta\rho_a$:
\begin{equation}
	\frac{d\rho_a}{dz} - \frac{3}{1+z}\rho_a = -\frac{\nu}{(1+z)H} \left[ \rho_a - \rho_a^{eq}(z=0)(1+z)^3 \right]
	\, .
	\label{ax6}
\end{equation}

Finally, the Friedmann equation governing the isotropic dynamics of the Universe in this scenario takes the form:
\begin{equation}
	H^2(z) = \frac{\chi}{3} \left[ \rho_b^0(1+z)^3 + \rho_a(z) + \rho_{\Lambda} \right]
	\, , 
	\label{ax7}
\end{equation}
where $\chi$ is the Einstein gravitational constant, $\rho_b^0$ is the present-day baryon density, and $\rho_{\Lambda}$ represents the constant vacuum energy density.

\subsection{Scalar field dynamics}\label{sec2a}
We now restate the previous kinetic approach in terms of classical field dynamics to link the microphysics of axions to the macroscopic representation required for high occupation numbers of the axion boson state. To this end, we decompose the classical axion field into a dominant equilibrium component, $\theta^{eq}(t)$, and a small perturbation, $\delta \theta$. For the equilibrium component, the analysis presented in Sec.~\ref{sec1} remains fully applicable. 

For the independent perturbation, we consider its dynamics in the limit $\nu \gg H_0$, consistent with the emergence of an efficient BGK term in the very late Universe. Accordingly, $\delta \theta$ obeys the following field equation:
\begin{equation}
	\ddot{\delta \theta} + (3H - \nu) \dot{\delta \theta} + m_a^2 \delta \theta = 0
	\, , 
	\label{ax8}
\end{equation}
which admits the leading-order solution:
\begin{equation}
	\delta \theta = \frac{\delta \theta_0\,e^{\nu t/2}}{a^{3/2}} \sin(m_a t + \delta \theta^*)
	\, , 
	\label{ax9}
\end{equation}
where $\delta \theta_0$ and $\delta \theta^*$ are integration constants. We now define the perturbative energy density as $\delta \rho_a \equiv \frac{\sigma_{SSB}^2}{2}(\dot{\delta \theta}^2 + m_a^2 \delta \theta^2)$. From Eq.~(\ref{ax8}), we derive the following evolution equation:
\begin{equation}
	\dot{\delta \rho}_a = -(3H - \nu) \sigma_{SSB}^2 \langle \dot{\delta \theta}^2 \rangle \simeq -(3H - \nu) \delta \rho_a
	\, , 
	\label{ax10}
\end{equation}
where the last equality is obtained by substituting the solution from Eq.~(\ref{ax9}) and averaging over many oscillations of the scalar field perturbation. Consequently, as long as the scalar field components $\theta^{eq}$ and $\delta \theta$ evolve independently, we recover the macroscopic behavior of Eq.~(\ref{ax4}) within the field-theoretical formulation as well.

We stress that the field‑theoretic construction above reproduces the macroscopic evolution derived from the kinetic BGK‑like ansatz but does not constitute an independent microscopic derivation of the collision parameter $\nu$. In other words, this section provides a mathematically consistent reformulation of the phenomenological source term introduced in \eqref{ax2} rather than a microphysical origin for it. The physical interpretation and microphysical generation of $\nu$ remain unspecified.

\section{Modified $\Lambda$CDM model: Data analysis, Methodology and Results}\label{sec3}
We introduce the dimensionless density parameters $\Omega_x = \chi \rho_x / 3H_0^2$ (where $x = a, b, \Lambda, \text{eq}$), allowing Eqs.~(\ref{ax6}) and (\ref{ax7}) to be expressed as:
\begin{equation}
	\frac{d\Omega_a}{dz} = \frac{3\Omega_a}{1+z} - \frac{\nu_0}{(1+z)E(z)} \left[ \Omega_a - \Omega_a^{eq}(1+z)^3 \right]
	\, , 
	\label{axf1}
\end{equation}
and
\begin{equation}
	E^2(z) \equiv \left( \frac{H}{H_0} \right)^2 = \Omega_b^0(1+z)^3 + \Omega_a(z) + 1 - \Omega_m^0
	\, , 
	\label{axf2}
\end{equation}
respectively. Here, we have defined the normalized collision frequency $\nu_0 \equiv \nu/H_0$ and the total present-day matter density $\Omega_m^0 \equiv \Omega_b^0 + \Omega_a^0$. The vacuum energy contribution is fixed as $\Omega_{\Lambda} \equiv 1 - \Omega_m^0$ to satisfy the consistency condition $E(z=0)=1$.

This model is characterized by five independent parameters: $H_0$, $\Omega_m^0$, $\Omega_b^0$, $\Omega_a^{eq}$, and $\nu_0$. For $\nu_0 \gg 1$, the evolution of $\Omega_a$ is driven from its present-day value toward the equilibrium trajectory, $\Omega_a^{eq}(1+z)^3$. Once this equilibrium state is reached at a characteristic redshift $z_{eq}$, the axion density retains the standard scaling $\Omega_a \propto (1+z)^3$ throughout the earlier stages of the Universe. 

To evaluate the statistical viability of the present axion model, we perform a robust model comparison against the $\Lambda$CDM baseline using the Bayesian evidence, $E(D|M)$. This is derived from Bayes’ theorem:
\begin{equation}
    P(u|D,M)= \frac{L(D|u,M)P(u|M)}{E(D|M)},
\end{equation}
where $u$, $M$, and $D$ represent the parameters, model, and data, respectively. Parameter inference is conducted by maximizing the likelihood $L \propto e^{-\chi^2/2}$, where the total $\chi^2$ for the combined datasets is:
\begin{equation}
    \chi^2_{\tt data} = (d_{m}-d_{\tt data})^{T}\, C^{-1}_{\tt data}\, (d_{m}-d_{\tt data}).
\end{equation}
The comparison between our proposed model ($M_i$) and the reference $\Lambda$CDM is quantified by the Bayes factor $B_{i, \Lambda\text{CDM}} \equiv E(D|M_i)/E(D|\Lambda\text{CDM})$. We report the results in terms of the log-evidence:
\begin{equation}
    \ln B_{i, \Lambda\text{CDM}} = \ln E(D|M_i) - \ln E(D|\Lambda\text{CDM}).
\end{equation}
Following the revised Jeffreys scale \cite{Trotta:2008qt}, a value of $\ln B_{i, \Lambda\text{CDM}} > 0$ indicates a preference for the axion model. The evidence is interpreted as: inconclusive if $|\ln B| < 1$, weak if $1 < |\ln B| < 2.5$, moderate if $2.5 < |\ln B| < 5$, and strong if $|\ln B| > 5$.
In addition to the Bayesian evidence, we assess the goodness-of-fit for each model by reporting the difference in the maximum log-likelihood relative to the baseline:
\begin{equation}
    -2\Delta\ln \mathcal{L}_{\rm max} = -2 \left( \ln \mathcal{L}_{\rm max}^{\Lambda\text{CDM}} - \ln \mathcal{L}_{\rm max}^{M_i} \right) = \chi^2_{\min, \Lambda\text{CDM}} - \chi^2_{\min, M_i}.
\end{equation}
In our results, a positive value of $-2\Delta\ln \mathcal{L}_{\rm max}$ indicates an improvement in data fitness for the non-equilibrium axion model over $\Lambda$CDM, whereas a negative value signifies a statistically inferior fit. This metric allows us to distinguish between models that provide a better mathematical description of the data points and those that are favored by the Bayesian evidence after accounting for the Occam's razor penalty associated with additional parameters ($\nu_0, \Omega_a^{eq}$).

For the $\Lambda$CDM baseline, we consider the physical baryon density $\Omega_b h^2 \in [0.02, 0.025]$, the matter density $\Omega_m \in [0.1, 0.9]$, and the Hubble constant $H_0 \in [40, 90]$. Our model introduces the two fundamental parameters derived above: the normalized collision frequency $\nu_0 \in [0.5, 20]$ and the equilibrium density $\Omega_a^{eq} \in [0.1, 0.6]$. Given our focus on the late-Universe dynamics and the assumption of a flat spatial geometry, radiation and curvature contributions are taken as negligible ($\Omega_r \approx 0, \Omega_k = 0$). Furthermore, we impose a Gaussian prior on the baryon density consistent with Big Bang Nucleosynthesis (BBN) constraints \cite{Schoneberg:2024ifp}, which drives the posterior of $\Omega_b h^2$ toward $0.02218 \pm 0.00055$.

Numerical parameter estimation is performed using the \texttt{SimpleMC} code~\cite{simplemc}, which employs the \texttt{dynesty} library~\cite{speagle2020dynesty} to implement the Nested Sampling algorithm~\cite{2004AIPC..735..395S}. This allows for the simultaneous extraction of posterior distributions and the Bayesian evidence required for our comparative analysis.

The analysis incorporates the following datasets: the \texttt{CC} catalog, consisting of 15 Cosmic Chronometers and their covariance matrix~\cite{Moresco:2020fbm}; the Pantheon+ (\texttt{PP}) sample of Type Ia Supernovae (SNe Ia)~\cite{Scolnic:2021amr,Brout:2022vxf}, used both independently and with the SH0ES Cepheid calibration (\texttt{PP\&SH0ES})~\cite{Riess:2021jrx}; the \texttt{DES-Dovekie} (\texttt{DD}) sample \cite{Popovic:2025glk, DES:2025sig}, which addresses photometric systematics in the DESY5 release\footnote{We note that while the ``Dovekie'' analysis identifies biases across various catalogs, recalibrated likelihoods for Union3 and Pantheon+ remain restricted by public availability.}; the Union3 (\texttt{U3}) compilation of 2087 SNe~\cite{Rubin:2023ovl}; and the latest Baryon Acoustic Oscillation (\texttt{BAO}) measurements from DESI DR2~\cite{DESI:2025zgx,DESI:2025fii,DESI:2025qqy}, providing constraints on the transverse ($D_M/r_d$), radial ($D_H/r_d$), and volume-averaged ($D_V/r_d$) distances.

\subsection{Results}\label{sec3a}
\begin{table*}[t]
\caption{The table summarizes the mean values and the standard deviations.}
\footnotesize
\scalebox{1.0}{%
\begin{tabular}{cccccccc} 
\cline{1-8}\noalign{\smallskip}
\vspace{0.15cm}
Model & Datasets & $H_0$ &  $\Omega_{m}$  & $\Omega_a^{eq}$ & $\nu_0$
& $\ln B_{i, \Lambda \text{CDM}}$  &  $-2\Delta\ln \mathcal{L_{\rm max}}$ \\
\hline
\hline
\vspace{0.15cm}
$\Lambda$CDM & CC+BAO+U3 &  $68.52 \pm 0.55$ & $0.305 \pm0.008$ & $-$ & $-$ & $-$  &  $-$  \\
\vspace{0.15cm}
Our model    & CC+BAO+U3 &  $69.23 \pm 2.52$ & $0.319 \pm0.024$ & $0.285\pm0.073$ & $1.728 \pm 0.418$ & $-3.56$  & $0.637$ \\
\hline
\vspace{0.15cm}
$\Lambda$CDM & CC+BAO+DD &  $68.52 \pm 0.52$ & $0.301 \pm0.008$ & $-$ & $-$ & $-$  &  $-$  \\
\vspace{0.15cm}
Our model    & CC+BAO+DD &  $68.93 \pm 2.42$ & $0.309 \pm0.023$ & $0.258 \pm 0.055$ & $1.887 \pm 0.381$ & $-4.99$  &  $0.808$ \\
\hline
\vspace{0.15cm}
$\Lambda$CDM & CC+BAO+PP &  $68.55 \pm 0.53$ & $0.305 \pm0.008$ & $-$ & $-$ & $-$  &  $-$  \\
\vspace{0.15cm}
Our model    & CC+BAO+PP &  $69.11 \pm 2.49$ & $0.316 \pm0.023$ & $0.276 \pm 0.068$ & $1.782 \pm 0.423$ & $-3.63$  & $0.354$ \\
\hline
\vspace{0.15cm}
$\Lambda$CDM & CC+BAO+PP\&SH0ES &  $69.74 \pm 0.47$ & $0.309 \pm0.008$ & $-$ & $-$ & $-$  &  $-$  \\
\vspace{0.15cm}
Our model    & CC+BAO+PP\&SH0ES &  $72.95 \pm 0.89$ & $0.351 \pm0.013$ & $0.229 \pm 0.056$ & $1.999 \pm 0.408$ & $5.32$  & $19.561$ \\
\hline
\hline
\end{tabular}}
\label{table_params}
\end{table*}

\begin{figure}
    \centering
    \includegraphics[width=0.45\linewidth]{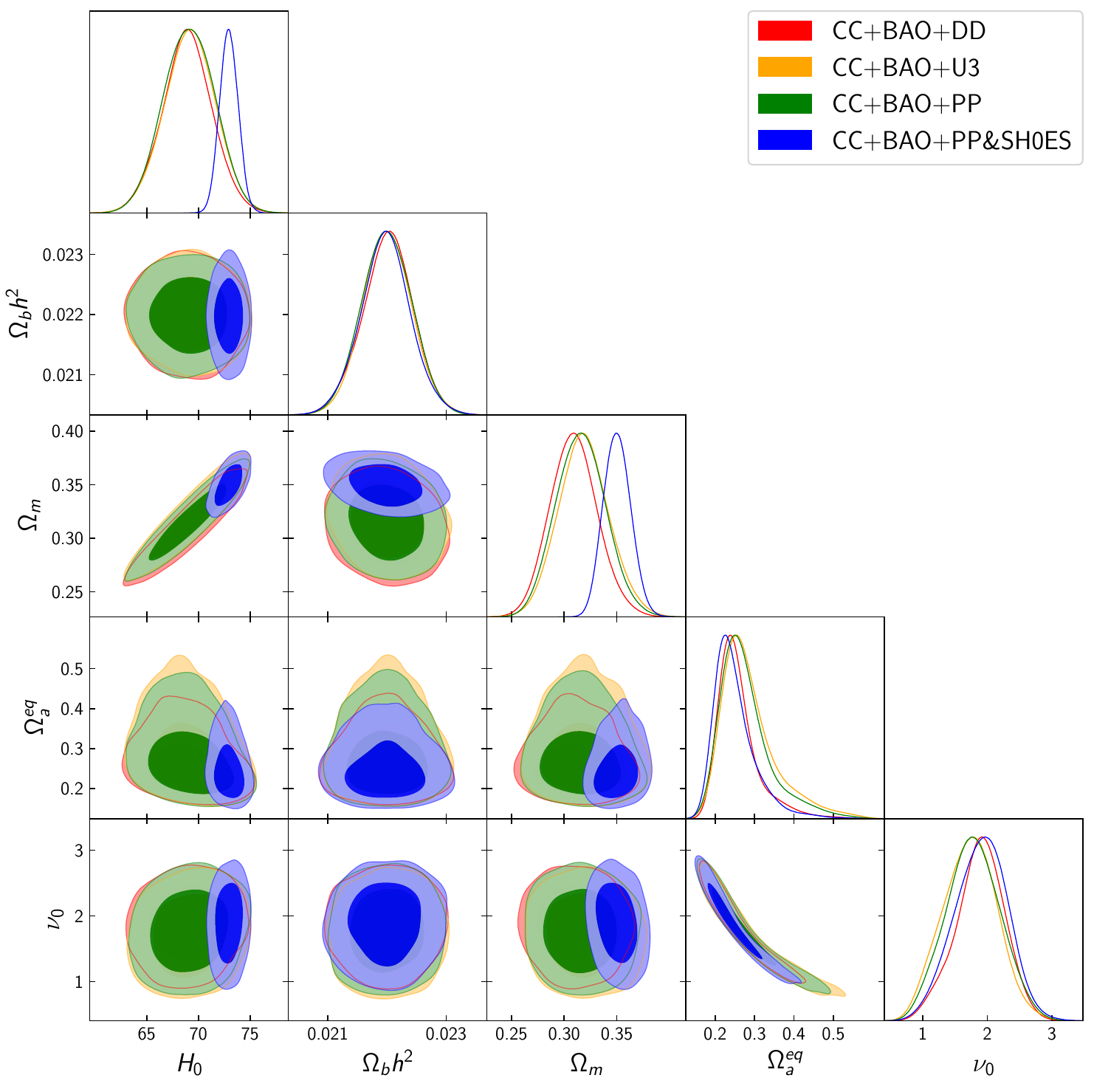}
    \includegraphics[width=0.45\linewidth]{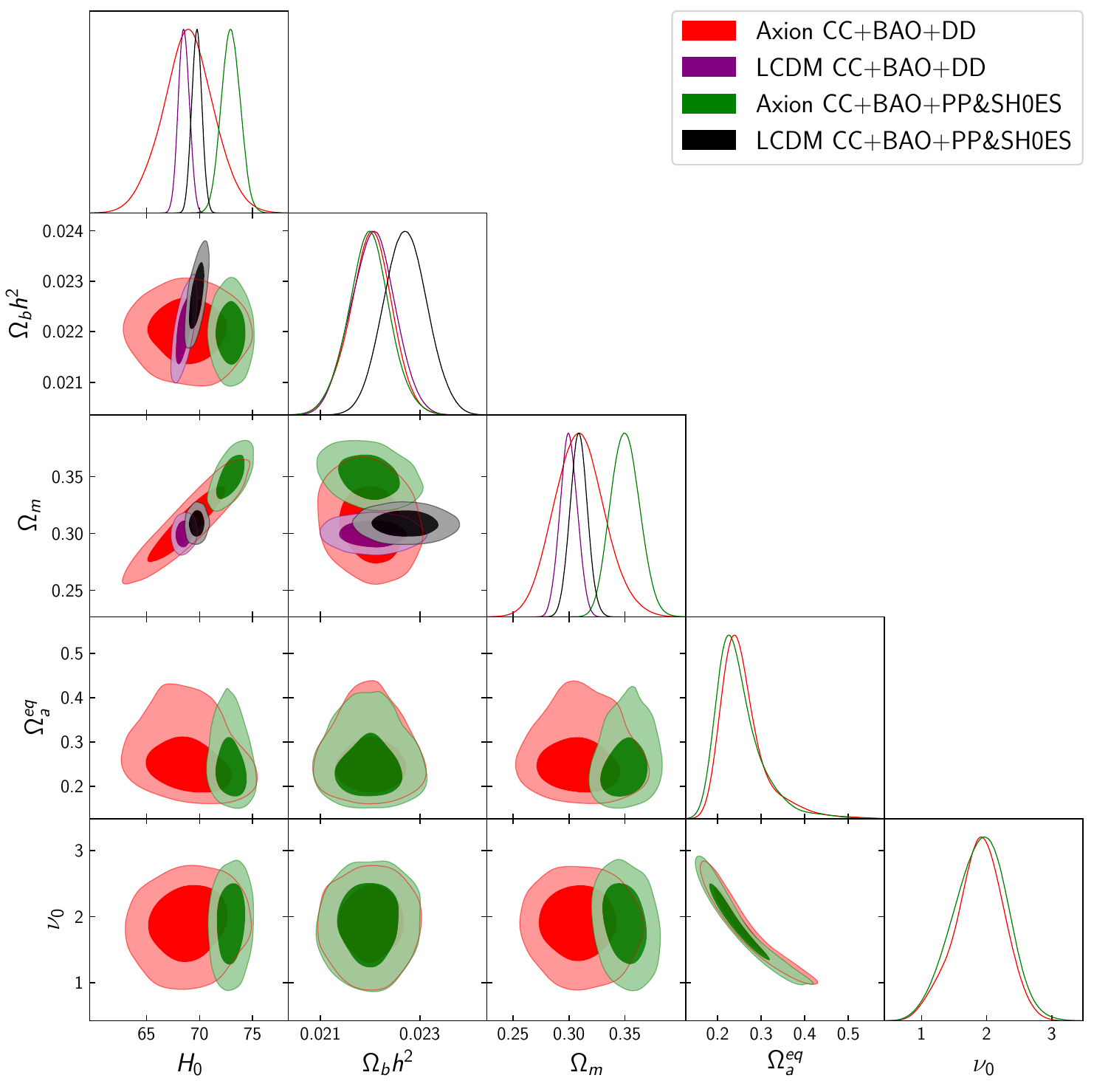}
    \caption{Marginal posteriors (1D and 2D) for the runs of: the Axion model with every data combination (left) and for the Axion model and $\Lambda$CDM for the dataset CC+BAO+DD and CC+BAO+PP\&SH0ES (right).}
    \label{fig:triangleplots}
\end{figure}

In Table~\ref{table_params}, we report the inferred mean values and $1\sigma$ uncertainties for the cosmological parameters, while Figure~\ref{fig:triangleplots} displays the marginalized 1D and 2D posterior distributions for each dataset combination.

As shown in the left panel of Figure~\ref{fig:triangleplots}, the posterior distributions exhibit significant overlap across most dataset combinations, suggesting a high degree of consistency between the late-time probes. The primary exception is the case incorporating the SH0ES calibration, which, as expected, significantly shifts the constraints. This shift is most pronounced in the $H_0$ and $\Omega_m$ planes; however, it is notable that the model-specific parameters, $\Omega_a^{eq}$ and $\nu_0$, are slightly affected by this inclusion. This is not obvious from the posterior distributions, but in Table~\ref{table_params} we see a shift in their mean values.

A comparison with the standard $\Lambda$CDM paradigm in the right panel of Figure~\ref{fig:triangleplots} reveals that our Axion model grants both $\Omega_m$ and $H_0$ considerably more freedom than the fiducial model. This increased flexibility is also evident in the numerical results: the $1\sigma$ uncertainty for $H_0$ is nearly five times larger than that of the standard model, while the uncertainty for $\Omega_m$ is approximately three times larger.

The model-comparison metrics, $\ln B_{\Lambda \text{CDM},i}$ and $-2\Delta\ln \mathcal{L}_{\rm max}$, reveal an interesting trend regarding the model's overall performance. For most dataset combinations, the improvements in data fitness ($-2\Delta\ln \mathcal{L_{\rm max}}$) are insufficient to offset the statistical penalty for the two additional parameters. This results in a Bayesian evidence ($\ln B_{\Lambda \text{CDM},i}$) that favors $\Lambda$CDM. However, the Axion model shines when the SH0ES calibration is implemented. In this regime, the model achieves a substantial advantage in fitness, yielding a Bayes' factor of approximately $-5$, indicating strong evidence in favor of our model.

A closer inspection of the model-specific parameters in Table~\ref{table_params} reveals a robust preference for a normalized collision frequency $\nu_0 \approx 2$ across all analyzed dataset combinations. This stability suggests that the onset of the non-equilibrium axion regime is a consistent feature of the model's interaction with late-time data. Furthermore, we observe a significant shift in the equilibrium density parameter $\Omega_a^{eq}$ when the SH0ES calibration is included. In this regime, $\Omega_a^{eq}$ decreases to $0.229 \pm 0.056$, while the matter density $\Omega_m$ increases to $0.351 \pm 0.013$. This transition characterizes the model's ability to accommodate a higher local expansion rate; by driving the system further from its equilibrium trajectory, the axion-creation mechanism generates an effective phantom contribution that raises $H_0$ to $72.95 \pm 0.89$, effectively resolving the tension with local measurements while maintaining a competitive fit to BAO and CC data.

\begin{figure}
    \centering
    \includegraphics[width=0.9\linewidth]{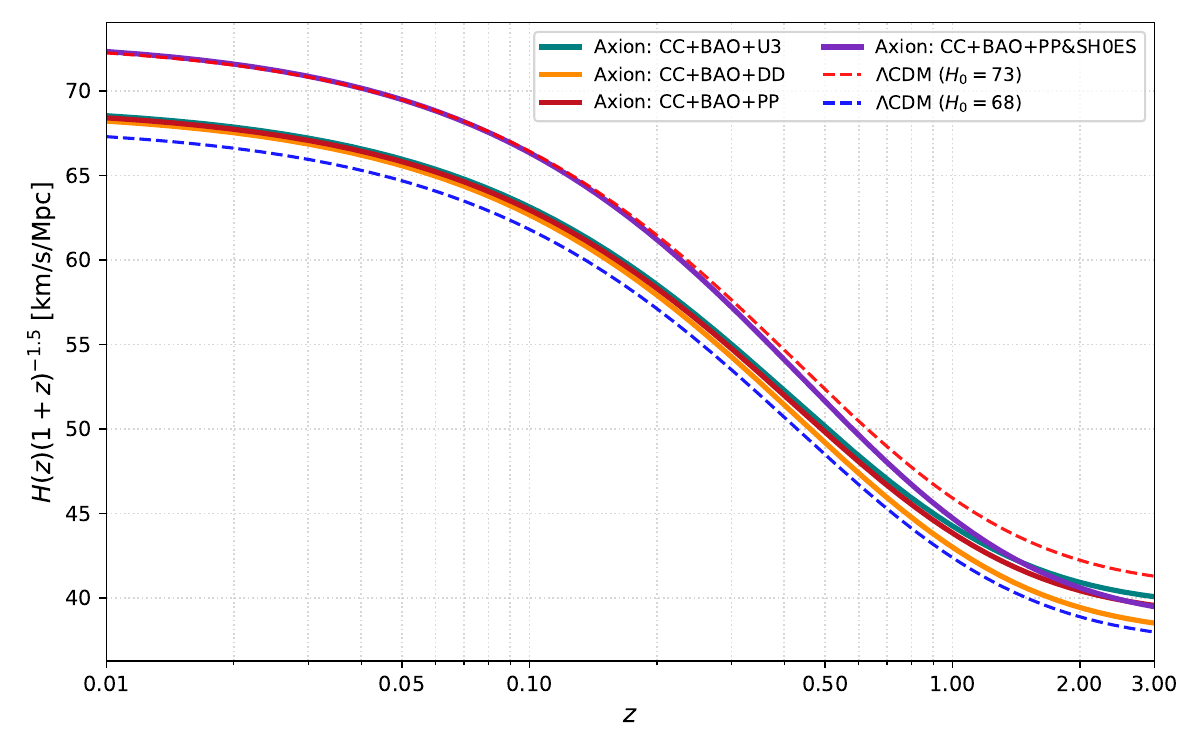}
    \caption{Evolution of the reduced Hubble parameter $H(z)/(1+z)^{1.5}$ for the Axion model across different data combinations, compared against two baseline $\Lambda$CDM models with $H_0 = 68~{\rm km\,s^{-1}\,Mpc^{-1}}$ and $H_0 = 73~{\rm km\,s^{-1}\,Mpc^{-1}}$. The axion-creation dynamics successfully reconcile these disparate expansion behaviors only when the SH0ES local prior is explicitly implemented. This demonstrates that while the model possesses the structural flexibility to mitigate the $H_0$ tension, this tracking behavior must be actively driven by the local calibration data.}
    \label{fig:hz_comparison}
\end{figure}

Nevertheless, these results warrant a cautious interpretation. The Axion model's success in the SH0ES-inclusive regime stems from its unique capability to reconcile the high local $H_0$ values with the lower $H_0$ values preferred by the DESI BAO data. This mechanism is explicitly visualized in Figure~\ref{fig:hz_comparison}, where the CC+BAO+PP\&SH0ES case utilizes the Axion dynamics to fit the SH0ES value at very low redshifts ($z < 0.1$). In contrast, the other three cases do not exhibit this behavior, as they are not driven by a requirement to match a high local expansion rate.

\section{Conclusions}\label{sec4}
We have developed a modified late Universe cosmology based on the premise that dark matter consists of axions, a small fraction of which is produced through some interactions that drive the system away from thermodynamic equilibrium. The underlying physical framework has been formulated through both a kinetic theory description and a classical field theory approach. These two complementary representations of the fundamental concept—that axions remain partially out of equilibrium during late-time cosmic expansion—lead to an identical modified evolution equation for the dark matter energy density, which serves as the foundation of our dynamical analysis. Our formulation yields a modified Hubble parameter that differs from the $\Lambda$CDM prediction, with the latter recovered only as an asymptotic limit. 

We then conducted a comprehensive comparison of our late Universe model against observational data from local probes, including SNe Ia, BAO, and CC. As discussed in the preceding section, our model exhibits a distinctive characteristic: without the SH0ES calibration, it performs statistically comparably to $\Lambda$CDM and fails to meaningfully alleviate the Hubble tension. However, when the SH0ES calibration is incorporated, the situation changes dramatically. In this case, our model is statistically favored over $\Lambda$CDM, with its associated Bayes factor approximately five units lower than that of $\Lambda$CDM, and notably, the inferred Hubble constant approaches 73 km s$^{-1}$ Mpc$^{-1}$. In other words, the present model appears capable of accommodating a higher Hubble constant value while simultaneously providing a robust fit to both SH0ES-calibrated SNe Ia and DESI BAO data. Consequently, our formulation has the significant advantage of successfully addressing the tension in the Hubble constant between SH0ES SNe Ia and DESI BAO measurements, arguably the central issue underlying the so-called Hubble tension.

We conclude by proposing a physical interpretation for the processes driving the axion distribution away from equilibrium: specifically, we draw on existing literature to identify mechanisms that could generate the deviations from $\Lambda$CDM discussed here. \eqref{ax2} models scattering and decay processes producing axions as out-of-equilibrium contributions, analogous to the treatment of hot axions in the early Universe in \cite{DEramo_2018}. The primary source of axion production in the late Universe is the decay of topological defects (predominantly cosmic strings) formed during the post-inflationary Peccei-Quinn phase transition through the Kibble mechanism \cite{KIBBLE1980183}. The magnitude of this contribution to cold axion dark matter remains contentious, as discussed in \cite{10.21468/SciPostPhys.10.2.050,Dine:2020pds,Buschmann:2021sdq}. To our knowledge, treating cosmic string axion production as an out-of-equilibrium contribution represents a novel approach. The prospect of using cosmological observations to constrain axion physics constitutes an intriguing avenue opened by this analysis.   

Our proposal assumes a specific chain of hypotheses: (i) the existence of axions comprising a fraction of the dark matter; (ii) Peccei–Quinn symmetry breaking after inflation producing a network of topological defects; and (iii) an efficient, macroscopically significant injection of out‑of‑equilibrium axions occurring in the very late Universe ($z \lesssim 1$). Each ingredient is individually plausible in some contexts, but their joint realization can be restrictive.

We stress that there is no established, widely accepted microphysical model in which a post‑inflationary network of axion strings releases a macroscopically significant, cold axion population predominantly at very late times. State‑of‑the‑art field‑theory and lattice studies generally find that string networks approach a scaling regime and emit most of their energy well before recent epochs, and that the axions radiated from strings tend to be relativistic (contributing as hot/warm components) or else contribute to the cold dark‑matter abundance at early times rather than as a late injection \cite{10.21468/SciPostPhys.10.2.050,Buschmann:2021sdq}. Achieving a late, cold, and macroscopically large release therefore requires non‑standard or highly tuned model ingredients (for example, suppression of early decay channels, a specially engineered late destabilization or coupling to an additional sector, or mechanisms that efficiently cool emitted axions), all of which demand explicit model building and dedicated simulation to assess viability. Recent analyses that explore conditions for “early vs late” network behaviours illustrate that such regimes can appear only in restricted regions of parameter space and typically require additional assumptions about the radial mode or other sector couplings \cite{2024JHEP...02..223G}.

Given the chain of restrictive assumptions underpinning the scenario studied, we present these results as a phenomenological exploration. The late‑time axion‑creation ansatz is speculative and requires additional microphysical input to be embedded in a complete particle‑physics model. While our statistical analysis shows that such a phenomenological modification can accommodate SH0ES‑inclusive local priors together with DESI BAO, this should not be interpreted as proof that the specific microphysics considered occurs in nature. Future work should investigate concrete microphysical realizations that can naturally produce the required timing and kinematics of the axion injection and test compatibility with detailed simulations of string evolution and axion emission.

Reconciling our ansatz with the detailed outcomes of string dynamics or demonstrating an explicit microphysical mechanism that yields late, cold injections is left to future work. For clarity, the model presented should therefore be read as an exploratory, phenomenological framework that parametrizes the possible cosmological impact of late out‑of‑equilibrium axions, rather than as a definitive, microphysically derived solution to the $H_0$ tension.

\section*{ACKNOWLEDGEMENTS}

L.A.E.\ acknowledges support from T\"{U}B\.{I}TAK through postdoctoral researcher fellowships associated with Grant No.~124N627.


\end{document}